\documentclass[11pt]{article}
\usepackage{jheppub}

\usepackage{graphics}
\usepackage{amssymb}
\usepackage{amsfonts}
\usepackage{amsbsy}
\usepackage{amsmath}

\usepackage{multirow}
\usepackage{mathtools}
\usepackage[table]{xcolor}

\usepackage{latexsym}
\usepackage{mathrsfs}

\usepackage[vcentermath]{youngtab}
\usepackage{graphicx}
\usepackage{amscd}
\usepackage{epstopdf}
\usepackage{amsthm}
\usepackage{enumerate}

\newtheorem{theorem}{Theorem}[section]

\newtheorem{proposition}[theorem]{Proposition}

\newcommand{\cO}{\mathcal{O}}

\newcommand{\cS}{\mathcal{S}}

\newcommand{\cX}{\mathcal{X}}

\newcommand{\bP}{\mathbb{P}}

\def\ov{\over}

\def\eq#1{(\ref{#1})}

\def \ra {\rightarrow}

\newcommand{\be}{\begin{equation}}
\newcommand{\ee}{\end{equation}}
\newcommand{\bea}{\begin{equation} \begin{aligned}}
\newcommand{\eea}{\end{aligned} \end{equation}}

\newcommand{\bln}{\begin{align}}
\newcommand{\eln}{\end{align}}
\newcommand{\bst}{\begin{split}}
\newcommand{\est}{\end{split}}
\newcommand{\bi}{\begin{itemize}}
\newcommand{\ei}{\end{itemize}}
\newcommand{\ben}{\begin{enumerate}}
\newcommand{\een}{\end{enumerate}}

\newcommand{\bprop}{\begin{proposition}}
\newcommand{\eprop}{\end{proposition}}

\newcommand{\tx}{\widetilde{x}}
\newcommand{\ty}{\widetilde{y}}

\title{Tall sections from non-minimal transformations}

\author[1]{David R. Morrison}
\author[2]{and Daniel S. Park}

\affiliation[1]{Departments of Mathematics and Physics \\
University of California, Santa Barbara\\
Santa Barbara, CA 93106, USA}
\affiliation[2]{NHETC and Department of Physics and Astronomy\\
Rutgers University, Piscataway, NJ 08855-0849, USA}

\emailAdd{drm {\rm at} math.ucsb.edu}
\emailAdd{dspark {\rm at} physics.rutgers.edu}

\abstract{In previous work, we have shown that elliptic fibrations with two sections, or Mordell-Weil rank one, can always be mapped birationally to a Weierstrass model of a certain form, namely, the Jacobian of a $\mathbb{P}^{112}$ model. Most constructions of elliptically fibered Calabi-Yau manifolds with two sections have been carried out assuming that the image of this birational map was a ``minimal'' Weierstrass model. In this paper, we show that for some elliptically fibered Calabi-Yau manifolds with Mordell-Weil rank-one,
the Jacobian of the $\mathbb{P}^{112}$ model is not minimal.  Said another
way, starting from a Calabi--Yau Weierstrass model, the total space must
be blown up (thereby destroying the ``Calabi--Yau'' property) in order
to embed the model into $\mathbb{P}^{112}$.
In particular, we show that the elliptic fibrations studied recently by Klevers and Taylor fall into this class of models.}

\begin{document}

\setcounter{tocdepth}{2}

\maketitle

%%%%%%%%%%%%%%%%%%%%%%%%%%%%%%%%%%%%%%%%%%%%
%
%              Body of paper
%
%%%%%%%%%%%%%%%%%%%%%%%%%%%%%%%%%%%%%%%%%%%%

\section{Introduction}

F-theory \cite{Vafa,MV1,MV2} has been a powerful tool for engineering a wide range of string vacua in diverse dimensions. While F-theory vacua that have already been found are extremely rich in the range of physical data---such as gauge groups or matter representations---that it can have, exploration of F-theory vacua are still well underway with expectations of more possibilities being in store.

A geometric F-theory compactification on an elliptically fibered Calabi-Yau (CY) manifold $\cX$ over the base $B$ can be thought of as a type IIB compactification on $B$ with a varying axio-dilaton profile. The value of the axio-dilaton is given by the complex structure of the elliptic fiber over each point. This information is summarized by a Weierstrass equation of the elliptic fibration
\be\label{W}
y^2 = x^3 + fx + g \,,
\ee
where $x$, $y$, $f$ and $g$ are sections of the bundles $-2K$, $-3K$, $-4K$ and $-6K$, respectively. Here, $K$ is the canonical bundle of the base manifold $B$.

Recently, there has been an interesting development in constructing F-theory models with abelian gauge symmetry with novel matter representations. In \cite{KT}, the authors study a family of F-theory vacua, first discovered in \cite{Klevers:2014bqa}, whose gauge algebra is given by $\mathfrak{u(1)}$ and has matter with charge three-times of the minimal charge.

For $\cX$ with complex dimension $\geq 3$, in order for an F-theory vacua compactified on $\cX$ to have gauge algebra $\mathfrak{u(1)}$, the elliptic fibration must have two sections. In other words, the elliptic fibration must have Mordell-Weil (MW) rank-one. In \cite{MP}, it was shown that an elliptically fibered manifold with two sections must be birationally equivalent to a Weierstrass model with a particular structure, that is that there must exist some sections $b$ and $c_i$ of appropriate line bundles such that the coefficients of the Weierstrass model are given by:
\be\label{MP}
{y'}^2 = {x'}^3 + \left( c_1 c_3-b^2 c_0 -{c_2^2 \ov 3}\right)x'
+\left( c_0 c_3^2 -{1 \ov 3} c_1 c_2 c_3 +{2 \ov 27} c_2^3
-{2 \ov 3} b^2 c_0 c_2\ +{b^2 c_1^2 \ov 4} \right) \,.
\ee
(This structure is obtained in \cite{MP} by first birationally embedding
the elliptic fibration into a $\mathbb{P}^{112}$ bundle over the base,
and then computing the
relative Jacobian of the resulting fibration.)
What is interesting about the models investigated in \cite{KT}, which we henceforth refer to as Klevers-Taylor (KT) models, is that their Weierstrass form deviates from this general form, i.e., there does not exist $b$ and $c_i$ such that $f$ and $g$ can be written as \eq{MP} in general.%
\footnote{This is to be contrasted with most constructions of Calabi-Yau manifolds with MW rank-one in the literature, a small sample of which is collected in the bibliography \cite{Mayrhofer:2012zy, Braun:2013yti, Cvetic:2013nia, Braun:2013nqa, Borchmann:2013hta, Antoniadis:2014jma, Kuntzler:2014ila, Esole:2014dea, Lawrie:2015hia, Krippendorf:2015kta}, where the Weierstrass model of the manifold takes the form \eq{MP}.}
An easy way to see that this is the case is to compute the height of the generator of the MW groups of the elliptic fibrations  $\cX \ra \bP^2$ in \cite{KT}. Some of the heights exceed the upper bound on the height that the generating section of a Calabi-Yau model, whose Weierstrass equation takes the form \eq{MP}, must satisfy. We thus arrive at an apparent contradiction.

The resolution of the problem is that the models of \cite{KT} are birationally equivalent to Weierstrass models of the form \eq{MP} by a {\it non-minimal} transformation, i.e., a map that does not preserve the canonical class of the manifold. More concretely, assuming that the Weierstrass form of a Klevers-Taylor model is given by equation \eq{W}, we can find a section $a \in \cO(D)$ for some effective line bundle $D$ such that the Weierstrass model
\be\label{prime}
{y'}^2 = {x'}^3 + f a^4 {x'} + g a^6
\ee
can be put into the form \eq{MP}, i.e., there exist $b$ and $c_i$ such that
\be\label{relation}
fa^4 = c_1 c_3-b^2 c_0 -{c_2^2 \ov 3} \,, \quad
g a^6 = c_0 c_3^2 -{1 \ov 3} c_1 c_2 c_3 +{2 \ov 27} c_2^3
-{2 \ov 3} b^2 c_0 c_2\ +{b^2 c_1^2 \ov 4} \,.
\ee
Note that this elliptic fibration is no longer Calabi-Yau, since $f a^4 \in \cO(-4K + 4D)$ and $g a^6 \in \cO(-6K+6D)$.

Thus we find that new classes of CY manifolds with MW rank-one, such as the KT models, may be generated via non-minimal maps from non-CY manifolds. Any such model can be obtained by the following steps in principle:
\begin{enumerate}
\item Begin with a non-CY Weierstrass model \eq{MP} where $x \in \cO(-2K+2D)$ and $y \in \cO(-3K+3D)$, or with its ``parent'' non-CY model in
$\mathbb{P}^{112}$, which uses the same sections $b$ and $c_i$ 
as coefficients.
\item Tune the coefficients of the model such that \eq{relation} holds for some $f \in \cO(-4K)$, $g \in \cO(-4K)$ and $a \in \cO(D)$. 
\item Obtain the CY manifold \eq{W} via the blowdown map%
\footnote{It is interesting to note that over the points of the base $B$ other than those on the divisor $a=0$, the elliptic fibers of the Weierstrass models \eq{W} and \eq{MP} have the same complex structure. To put it differently, away from the codimension-one locus $a=0$, the $j$-functions of the axio-dilaton of the models \eq{W} and \eq{MP} have the same value.}
\be
x' \mapsto a^2 x\,, \quad y' \mapsto a^3 y\,. 
\ee
Note that this last step 
is part of the usual procedure of obtaining a ``minimal''
Weierstrass model from a ``non-minimal'' one, and involves a blowdown
in the total space (see sections II.3 and III.3 in \cite{MR1078016},
eq.~(15) in \cite{instK3}, as well as
 eq.~(2.6) and footnote 10 in \cite{nongeometries}).  The blowdown
restores the Calabi--Yau property, since the holomorphic $n$-form
vanishes on the divisor which is being blown down.
\end{enumerate}
While the possibility of constructing more general classes of MW rank-one CY manifolds in this manner was hinted in \cite{MP}, the KT models are, to our knowledge, the first examples where this scenario is realized. It would be interesting to understand how to systematically implement these steps to find new CY fibrations with MW rank-one.

Arguably the most interesting aspect of models obtained this way is that their sections are ``taller," whose meaning we define in section \ref{s:background}, than those that are birationally equivalent to \eq{MP} by a minimal map. A generating section being tall has a direct physical consequence when the elliptic fibration is a CY threefold---the taller the generating section of the Mordell-Weil group, the bigger the maximum charge of the matter under the corresponding $\mathfrak{u(1)}$. As mentioned, the Klevers-Taylor models have matter with $\mathfrak{u(1)}$ charge 3, while the matter in models constructed in \cite{MP} all have charge $\leq 2$.

This paper is organized as follows. In section \ref{s:background}, we review some background material. We first review basic bounds on the height of the rational section of a Weierstrass model of the form \eq{MP}. We then show how such bounds get looser by allowing the model \eq{MP} to be non-CY, and discuss the property of CY models obtained by a non-minimal map. We also define a convenient metric to compare heights of rational sections, and discuss its physical relevance. We present our main result in section \ref{s:result} by explicitly identifying $b$, $c_i$ and $a$ by which the Weierstrass coefficients of the KT models can be constructed via equation \eq{relation}. We conclude with some remarks in section \ref{s:conclusion}. There are several unwieldy expressions that are collected in the appendix.

\section{Background} \label{s:background}

Let us review some salient points about elliptic fibrations with two sections. It was shown in \cite{MP} that any elliptically fibration over a base $B$ with MW rank-one, is birationally equivalent to the Weierstrass model $\cX'$ \eq{MP}
\be\nonumber
{y'}^2 = {x'}^3 + \left( c_1 c_3-b^2 c_0 -{c_2^2 \ov 3}\right)x'
+\left( c_0 c_3^2 -{1 \ov 3} c_1 c_2 c_3 +{2 \ov 27} c_2^3
-{2 \ov 3} b^2 c_0 c_2\ +{b^2 c_1^2 \ov 4} \right)
\ee
whose generating section is given by
\be\label{section}
[X',Y',Z'] = \left[ c_3^3 -{2 \ov 3} b^2 c_2 ,\,
-c_3^3 +b^2 c_2 c_3-{1 \ov 2} b^4 c_1,\,
b\right]
\ee
for some $b$ and $c_i$. We have introduced the projective coordinates $X'$, $Y'$ and $Z'$ for which
\be
x' = X'/{Z'}^2 \,, \quad
y' = Y'/{Z'}^3 \,,
\ee
to make the equations less cluttered. The entries for $X'$, $Y'$ and $Z'$ of \eq{section} are assumed to be mutually prime in the sense that there does not exist $p$ such that
\be
p^2 | X' \,, \quad
p^3 | Y' \,, \quad
p | Z' \,.
\ee
Given that $x'$ and $y'$ are sections of $\cO(-2K+2D)$ and $\cO(-3K+3D)$ for some effective $D$, we find that
\bea
~[c_0] &= -4K+4D - 2[b] \,, &
\qquad [c_1] &= -3K+3D - [b] \,, \\
~[c_2] &= -2K+2D \,, &
\qquad [c_3] &= -K+D + [b] \,.
\eea
where we use brackets $[b]$ to denote the line bundle of which $b$ is a section. We find that
\be\label{bound}
[b] \leq -2K + 2D \,,
\ee
where in discussing line bundles or divisors,
\be\label{eff}
D_1 \geq D_2
\ee
means that $D_1-D_2$ is either trivial or effective.

Let us now assume that there exist $a \in \cO(D)$, $f \in \cO(-4K)$ and $g \in \cO(-6K)$ such that \eq{relation} holds:
\be\nonumber
c_1 c_3-b^2 c_0 -{c_2^2 \ov 3} = a^4f\,, \quad
c_0 c_3^2 -{1 \ov 3} c_1 c_2 c_3 +{2 \ov 27} c_2^3
-{2 \ov 3} b^2 c_0 c_2\ +{b^2 c_1^2 \ov 4} = a^6 g \,.
\ee
For the purpose of this paper, we further assume that
\be\label{section map}
c_3^2 -{2 \ov 3} b^2 c_2 = a^2 \tx\,, \quad
-c_3^3 +b^2 c_2 c_3-{1 \ov 2} b^4 c_1 = a^3 \ty
\ee
for some $\tx$ and $\ty$.%
\footnote{It would be interesting to understand the consequences of relaxing this condition.}
By the birational transformation
\be\label{map}
X = X' \,, \quad
Y = Y' \,, \quad
Z \mapsto aZ' \,, \quad
x = X/Z^3 \,, \quad
y = Y/Z^2 \,,
\ee
we arrive at a Calabi-Yau fibration \eq{W}
\be\nonumber
y^2 = x^3 + f x +g \,,
\ee
with MW rank-one with generating rational section
\be\label{S}
S~: ~[X,Y,Z] = \left[ c_3^2 -{2 \ov 3} b^2 c_2 ,\,
-c_3^3 +b^2 c_2 c_3-{1 \ov 2} b^4 c_1,\,
ab\right] = [\tx, \ty, b]\,.
\ee
The transformation \eq{map} is non-minimal when $D$ is non-trivial, i.e., $D > 0$.

The height $h(S)$ of a rational section $S$ is a divisor in the base manifold $B$ obtained by first taking the image
$\sigma(S)$ of $S$ under the
Shioda map \cite{Shioda1,Shioda2,Intersection} 
and projecting its self-intersection down to the base $B$ \cite{MP,Intersection}:
\be
h(S) = -\pi(\sigma(S),\sigma(S)) \,.
\ee
When the base $B$ is a surface, and there are no additional gauge group factors, i.e., there are no codimension-one singularities beyond type $I_1$, $h(S)$ is given by \cite{MP}
\be
h(S) = -2K + 2[b] \,.
\ee
We thus see that $h(S)$ is bounded by
\be
h(S) \leq -6K + 4D \,.
\ee
Note that in the examples studied in \cite{MP}, it was assumed that the elliptic fibration \eq{MP} itself was Calabi-Yau. This meant that $h(S)$ has an upper-bound $-6K$. We are able to construct new Calabi-Yau threefolds with MW rank-one given $b$ and $c_i$ with $[b] > -2K$ which satisfy equations \eq{relation} and \eq{section map} for some $a$, $f$, $g$, $\tx$ and $\ty$.

The height of the generating section $S$ has the physical interpretation as the corresponding $\mathfrak{u(1)}$ anomaly coefficient when $B$ is a surface \cite{MP,Intersection}.%
\footnote{An similar analysis of the role of $h(S)$ when $B$ is complex three-dimensional can be found in \cite{Cvetic:2012xn}.}
The anomaly cancellation conditions then imply that
\be\label{anomaly}
-K \cdot h(S) = {1 \ov 6}\sum_I q_I^2 \,, \quad
h(S) \cdot h(S) = {1 \ov 3} \sum_I q_I^4
\ee
where $I$ labels the hypermultiplets charged under the $\mathfrak{u(1)}$ algebra. When the gauge algebra is given solely by $\mathfrak{u(1)}$, the charges in \eq{anomaly} are integrally quantized. Inspired by the anomaly equations, we define the additional measure of {\it tallness}
\be
t(S) \equiv {h(S)^2 \ov -2K \cdot h(S)}
\ee
associated to a section $S$. While the height $h(S)$ is only partially ordered with respect to the relation defined in \eq{eff}, $t(S)$ is ordered since it is a number. We say that a section $S_1$ is {\it taller than} $S_2$ when
\be
t(S_1) > t(S_2) \,.
\ee
By the anomaly equations, it is simple to see that
\be
t(S) = {\sum_I q_I^4 \ov \sum_I q_I^2} \leq \max_I q_I^2 \,.
\ee
Thus the square of the maximum charge of the $\mathfrak{u(1)}$ algebra corresponding to the section $S$ is bounded below by the tallness of $S$. In particular, when $t(S) >4$, there must be matter with charge at least $3$ in the spectrum of the F-theory compacitification.

Note that when $\cX'$ itself is Calabi-Yau, and there are no codimension-one singularities beyond $I_1$, $h(S) \leq -6 K$. The divisor $h(S)$ is a positive self-intersection curve, and for generic $b$ and $c_i$, it is irreducible. It follows that
\be
h(S)^2 \leq -6K \cdot h(S)
\ee
which implies that $t(S) \leq 3$. Hence, none of these models are forced to have matter with charge $\geq 3$. On the other hand, when $h(S)$ is not restricted to be bounded above by $-6K$, it can happen that $t(S) > 4$, forcing the existence of matter with charge at least $3$. To be concrete, let us consider the case with $B = \bP^2$ and $\cX$ is obtained by a non-minimal transformation \eq{map} from $\cX'$. In this case, the homology lattice is one-dimensional, and all the divisor classes are proportional to the hyperplane class $H$ with $H^2 =1$. For example, $K = -3H$. Then $t(S)$ is given by
\be
t(S) H = {h(S) \ov 6}
\ee
and when $h(S) > 24 H$, or equivalently, $[b] > 9 H$, matter with charge $\geq 3$ is forced upon the model. In order for this to happen, we must have $D \geq 2H$. In \cite{KT}, twelve elliptic fibrations over $\bP^2$ with charge $3$ matter are listed, five of them with $h(S) > 24 H$.

\section{Klevers-Taylor models} \label{s:result}

In this section, we find $a$, $b$ and $c_i$ that relate to the $f$ and $g$ of the Klevers-Taylor models \eq{W}
\be\nonumber
y^2 = x^3 + fx +g
\ee
by equation \eq{relation}. The coefficients $f$ and $g$ of the KT models are expressed in terms of the sections $s_i$ of the line bundles listed in table \ref{t:si}. These explicit expressions may be found in the appendix of \cite{KT}.

\begin{table}[t!]
\begin{center}
\begin{tabular}{c || c | c |  c | c |  c }
\hline
Section & \multicolumn{2}{c |}{$s_1$} &
$s_2$ & $s_3$ & $s_4$ \\ \hline
Line Bundle & \multicolumn{2}{c |}{$-3K-\cS_7 -\cS_9$} &
$-2K-\cS_9$ & $-K+\cS_7 -\cS_9$ &
{$2\cS_7-\cS_9$} \\ \hline\hline
Section & $s_5$
& $s_6$ & $s_7$ & $s_8$ & $s_9$ \\ \hline
Line Bundle & $-2K -\cS_7$ & $-K$ &
$\cS_7$ & $-K-\cS_7 +\cS_9$ & $\cS_9$
\\ \hline
\end{tabular}
\end{center}
\caption{The $s_i$ are the sections of line bundles listed in this table. The line bundles are expressed in terms of
$\cS_7$ and $\cS_9$ which must be such that all the line
bundles listed in the table are effective.}
\label{t:si}
\end{table} 

The equation \eq{relation} is very hard to solve in general. What makes the solution possible is an observation made in the conclusion of \cite{KT}, where it is pointed out that the generating section of the KT models are given by
\be\label{KT section}
[X, Y ,Z] = 
\left[ a_3^2 - {2 \ov 3} b a_2,\, -a_3^3 + b a_2 a_3 -{1 \ov 2} b^2 a_1,  \, b \right]
\ee
for $a_i$ that can be explicitly written in terms of $s_i$. The explicit formulae for $a_i$ and $b$ are given in appendix \ref{ap:expressions}. Since the section \eq{KT section} should be given by equation \eq{S}, we have the additional conditions
\bea\label{KT section map}
c_3^2 - {2 \ov 3} b^2 c_2 &= a^2 \left( a_3^2 - {2 \ov 3} b a_2 \right) \\
-c_3^3 + b^2 c_2 c_3 -{1 \ov 2} b^4 c_1 &= a^3
\left(  -a_3^3 + b a_2 a_3 -{1 \ov 2} b^2 a_1 \right)
\eea
which are key in obtaining the desired $a$, $b$ and $c_i$.

Before going further, we make a few assumptions, that turn out to be true for the Klevers-Taylor models. We first assume that $a$ is irreducible, and that $a$ does not divide $b$. We also assume that $c_3$ is not divisible by $a$. If so, it follows that $a^2 | c_2$, $a^3 | c_1$ and $a^4 | c_0$. We do not wish to consider this trivial case. We need to make one more technical assumption, which we explain shortly.

Let us start with the first equation of \eq{KT section map}. We can arrange
\be\label{1}
(c_3 - a a_3) (c_3 + a a_3) = {2 \ov 3} b (bc_2 -a_2 a^2) \,.
\ee
Without loss of generality, we find that
\be
c_3 - a a_3 = bd
\ee
for some $d$. We plug this into \eq{1} and get
\be\label{2}
b ({3 \ov 2} d^2 -c_2) = - a (3 a_3 d +a_2 a) = {3 \ov 2} b a \kappa
\ee
for some $\kappa$, where we have assumed that $b$ and $a$ do not have a common divisor. We then arrive at
\be\label{c2c3}
c_2 = {3 \ov 2} (-a \kappa + d^2) \,, \quad
c_3 = bd + a a_3 \,.
\ee
Meanwhile,
\be
a a_2 = {1 \ov 2} (3 b \kappa + 6 a_3 d)\,.
\ee
Now the right-hand-side of this equation must be divisible by $a$. Here we make the assumption that {\it each of the two terms} on the right-hand-side of this equation are divisible by $a$. This implies that 
\be
\kappa = a k
\ee 
for some $k$. Also, $a$ must divide $a_3 d$. $a$ being irreducible, it must either divide $a_3$ or $d$. When $a$ divides $d$, $c_3$ is divisible by $a$ due to equation \eq{c2c3}, contrary to our assumption. It thus follows that $a_3$ is divisible by $a$.

In the KT models, $a_3$ is irreducible for generic complex structure. In fact, $a_3$ is the $SU(2)$ divisor at the point of enhanced gauge symmetry \cite{KT}. Hence, $a_3$ and $a$ must agree up to a constant factor. This constant can be absorbed trivially into the definition of the birational map \eq{map}. We can therefore identify $a_3$ with $a$. Plugging this into the above equations, we arrive at
\be\label{ac}
a_2 = -{3 \ov 2}(bk+2d)\,,\quad
a_3 = a \,, \quad
c_2 = {3 \ov 2}(-a^2 k +d^2) \,, \quad
c_3 = bd + a^2 \,.
\ee
While we have explicit expressions for $a_i$ and $b$, the equations written above alone do not determine $k$ and $d$ uniquely. Note that once $d$ is determined, so is $k$.

Upon plugging what we have into the second equation of \eq{KT section map}, we arrive at
\be
b (bc_1 - d^3) = a^2 (a a_1 -3d^2 - 3 a^2 b k d) \,.
\ee
It follows that there exists some $M$ such that
\be
b (bc_1 - d^3) = a^2 (a a_1 -3d^2 - 3 a^2 b k d) = b a^2 M \,.
\ee
which implies that
\be\label{M}
bc_1 = a^2 M + d^3 \,, \quad
a a_1 -3d^2 = b(M + 3 dk)  \,.
\ee

Now using the explicit expressions of the coefficients $a_i$ of the KT model, we find that
\be\label{divisible}
a_1 a_3-{1 \ov 3} a_2^2  = bL
\ee
for a polynomial $L$ of $s_i$. Using the expression for $a_2$ and $a_3$, and equation \eq{M}, we find that
\be\label{ML}
M = L+{3 \ov 4}bk^2 \,.
\ee
Plugging this into the first equation of \eq{M}, we arrive at
\be\label{c1_main}
c_1 = {3 \ov 4} a^2 k^2 + {a^2 L + d^3 \ov b} \,.
\ee
By the first equation in \eq{ac}, and the fact that
\be
b\, | \, (a^2 L -{1 \ov 27} a_2^3) \,,
\ee
which is another convenient miracle, $c_1$ is guaranteed to be well-defined. The non-trivial condition on $k$ and $d$ comes from the fact that $c_1$, $c_2$ and $c_3$ as given by equations \eq{ac} and \eq{c1_main}, must be such that $c_0$, as can be computed from the first equation of equation \eq{relation}, is a polynomial of the sections $s_i$. That is,
\be\label{divisibleb2}
b^2 \, \Big| \, \left( a^4 f - c_1 c_3 + {c_2^2 \ov 3}\right)
\ee
where the expression for the coefficient $f$ for the KT models is written out explicitly in the appendix of \cite{KT}. For example, equation \eq{divisibleb2} is not satisfied by the solution $(k,d) = (0,-a_2/3)$ of the first equation of \eq{ac}.

We can find $k$ and $d$ with
\be\label{a2}
a_2 = -{3 \ov 2}(bk+2d)
\ee
such that $d$ has the minimal number of terms as a polynomial of the sections $s_i$. $d$ and $k$ are given by
\bea\label{d}
d=& -{1 \ov 2} s_4 s_6 s_8^3 + s_4 s_5 s_8^2 s_9 + {1 \ov 2} s_3 s_6 s_8^2 s_9 - 
 s_3 s_5 s_8 s_9^2\\
& - {1 \ov 2} s_2 s_6 s_8 s_9^2 + s_1 s_7 s_8 s_9^2 + s_2 s_5 s_9^3 -  {1 \ov 2} s_1 s_6 s_9^3
\eea
and
\bea\label{k}
k =& {1 \ov 12} (s_6^2-4 s_5 s_7+8 s_3 s_8-16 s_2 s_9) \,.
\eea
This $d$, $b$ satisfies equation \eq{divisible}!
%Equation \eq{d} turns out to be the correct expression for the desired  $d$. Let us elaborate.
Since we now know $d$, $k$ and $L$, $c_1$ is obtained via equation \eq{c1_main}, and $c_0$ is given by
\be\label{c0_main}
c_0 = {1 \ov b^2} \left( -a^4 f + c_1 c_3 - {c_2^2 \ov 3}\right) \,.
\ee
By the way we have obtained $c_i$, it is not entirely clear that the second equation of \eq{relation} should hold, but it does.

Now \eq{d} and \eq{k} are not the only polynomials of $s_i$ such that \eq{d} has the minimal number of terms. There are, in fact nine such pairs of $(k,d)$ that are part of a two-parameter family of solutions of \eq{a2}:
\bea\label{dk}
k=& {1 \ov 12} s_6^2-{1 \ov 3} s_5 s_7+\ell_1 s_3 s_8 + \ell_2 s_2 s_9 \\
d=&
-{1 \ov 2} s_4 s_6 s_8^3 +s_4 s_5 s_8^2 s_9+  s_1 s_7 s_8 s_9^2
- {1 \ov 2} s_1 s_6 s_9^3\\
&+\left({1\ov 3} -{1\ov 2} \ell_1 \right) s_3 s_7 s_8^3 + 
  \left({1 \ov 6}  + {1 \ov 2} \ell_1 \right) s_3 s_6 s_8^2 s_9 
+\left(- {2 \ov 3}  -  {1 \ov 2} \ell_1 \right) s_3 s_5 s_8 s_9^2 \\
&+ 
 \left(-{2 \ov 3} - {1 \ov 2} \ell_2 \right) s_2 s_7 s_8^2 s_9
 + \left({1 \ov 6}  + {1 \ov 2} \ell_2 \right) s_2 s_6 s_8 s_9^2 + 
 \left({1 \ov 3}  - {1 \ov 2} \ell_2 \right) s_2 s_5 s_9^3 
\,.
\eea
In the solution given by \eq{d} and \eq{k}, $\ell_1$ and $\ell_2$ are set to
\be
\ell_1 = {2 \ov 3} \,, \quad
\ell_2 = -{4 \ov 3} \,.
\ee
For $d$ and $k$ of \eq{dk}, $c_0$ and $c_1$ defined by \eq{c0_main} and \eq{c1_main}
\be\label{c0c1}
c_0 = {1 \ov b^2} \left( -a^4 f + c_1 c_3 - {c_2^2 \ov 3}\right) \,, \qquad
c_1 = {3 \ov 4} a^2 k^2 + {a^2 L + d^3 \ov b} \,,
\ee
still remain polynomials of $s_i$, and solve the equations \eq{relation} and \eq{KT section map} for any values of $\ell_1$ and $\ell_2$. The explicit expression for $L$ is given in equation \eq{large L} of the appendix.

To summarize, we have found that the Klevers-Taylor models  \eq{W} are birationally equivalent to the models \eq{MP} by the map \eq{map}. The $x'$ and $y'$ of the Weierstrass model $\cX'$ given by \eq{MP} are sections of $\cO(-2K+2D)$ and $\cO(-3K+3D)$, where
\be
D = -3K-\cS_7+2\cS_9 \,.
\ee
This is because
\be
D = [a] = [a_3] \,.
\ee

There is a two-parameter family of solutions to the equations \eq{relation} and \eq{KT section map}. $b$, $c_2$ and $c_3$ of the model \eq{MP} for $\cX'$ can be expressed as
\be
b=s_7 s_8^2 - s_6 s_8 s_9 + s_5 s_9^2 \,, \quad
c_2 = {3 \ov 2}(-a^2 k +d^2) \,, \quad
c_3 = bd + a^2
\ee
where $a$ for which the mapping \eq{map} is defined, is given by
\be
a = a_3 = s_4 s_8^3 - s_3 s_8^2 s_9 + s_2 s_8 s_9^2 - s_1 s_9^3
\ee
and $d$ and $k$ are given by equations \eq{dk}:
\bea\nonumber
k=& {1 \ov 12} s_6^2-{1 \ov 3} s_5 s_7+\ell_1 s_3 s_8 + \ell_2 s_2 s_9 \\
d=&
-{1 \ov 2} s_4 s_6 s_8^3 +s_4 s_5 s_8^2 s_9+  s_1 s_7 s_8 s_9^2
- {1 \ov 2} s_1 s_6 s_9^3\\
&+\left({1\ov 3} -{1\ov 2} \ell_1 \right) s_3 s_7 s_8^3 + 
  \left({1 \ov 6}  + {1 \ov 2} \ell_1 \right) s_3 s_6 s_8^2 s_9 
+\left(- {2 \ov 3}  -  {1 \ov 2} \ell_1 \right) s_3 s_5 s_8 s_9^2 \\
&+ 
 \left(-{2 \ov 3} - {1 \ov 2} \ell_2 \right) s_2 s_7 s_8^2 s_9
 + \left({1 \ov 6}  + {1 \ov 2} \ell_2 \right) s_2 s_6 s_8 s_9^2 + 
 \left({1 \ov 3}  - {1 \ov 2} \ell_2 \right) s_2 s_5 s_9^3 
\,.
\eea
$c_0$ and $c_1$ are given by equations \eq{c0c1}:
\be\nonumber
c_0 = {1 \ov b^2} \left( -a^4 f + c_1 c_3 - {c_2^2 \ov 3}\right) \,, \qquad
c_1 = {3 \ov 4} a^2 k^2 + {a^2 L + d^3 \ov b} \,.
\ee
Explicit expressions for $c_0$ and $c_1$ when $\ell_1=2/3$ and $\ell_2 = -4/3$ are written out in equations \eq{c0} and \eq{c1} in the appendix. 

\section{Comments and future directions} \label{s:conclusion}

In this paper, we have shown that the Klevers-Taylor models can be obtained via non-minimal transformations from non-Calabi-Yau manifolds of the form \eq{MP} constructed in \cite{MP}. We conclude by commenting on our results and speculating on future directions to pursue.
\vspace*{0.05in}

\noindent
\textbf{Non-genericity of polynomials:} It is crucial that the polynomials $a_i$ and $b$ of $s_i$ involved in constructing the KT models satisfy intricate relations. For example, the relation \eq{divisible} played an important role in discovering the birational map from \eq{MP} to the KT models. Equation \eq{divisible} is possible because the ring of functions on $a_3$ is not a universal factorization domain (UFD), as pointed out in \cite{KT}.%
\footnote{The fact that the functions on $a_3$ is not a UFD does not come as a surprise. After all, $a_3$ becomes the $\mathfrak{su(2)}$ locus as the $\mathfrak{u(1)}$ gauge symmetry is un-Higgsed by taking $b \ra 0$. The charge-three hypermulitplets then organize themselves as part of the spin-$3/2$ representation of $\mathfrak{su(2)}$. If $a_3$ had been a generic smooth curve, such exotic representations of $\mathfrak{su(2)}$ cannot appear. See \cite{KT} and \cite{Cvetic:2015ioa} for more discussions.}
For this reason, we expect that constructing $\mathfrak{u(1)}$ theories exotic matter following the steps laid out in the introduction to be a very hard in general.
\vspace*{0.05in}

\noindent
\textbf{Generalization of KT models:}
It would be interesting to generalize the KT models in a meaningful way. Now the fact that the rational section of the KT models had the form \eq{KT section} was important in finding the rational map between them and the the Weierstrass models of the form \eq{MP}. A systematic understanding and generalization of the form \eq{KT section} would be desirable. Do the KT models saturate all Weierstrass models with sections of the form \eq{KT section}? Is there a nice generalization of the form \eq{KT section} which would lead to a new class of models? Understanding the answer to these questions would be imperative to either constructing new F-theory models with abelian gauge symmetry, or explaining why those constructions should not be possible.
\vspace*{0.05in}

\noindent
\textbf{Bounds on $\mathfrak{u(1)}$ charges and anomaly coefficients:}
An important physical question is what kind of theories are constructible in string theory. It would be desirable to gain a better understanding which values of physical observables such as abelian charges and anomaly coefficients of $\mathfrak{u(1)}$ gauge groups are allowed in F-theory models. A good first step would be to understand the constraints on the line bundle $D = [a]$ defined in the introduction in order for the Calabi-Yau fibration obtained by the birational map \eq{map} to have a smooth resolution.
\vspace*{0.05in}

\noindent
\textbf{Non-enhanceable $\mathfrak{u(1)}$s:} 
As noted in \cite{Morrison:2014era}, the $\mathfrak{u(1)}$ gauge symmetry for any F-theory model with the Weierstrass form \eq{MP} can be enhanced to a non-abelian gauge symmetry by taking $b \ra 0$, unless a codimension-one $(4,6,12)$ singularity is induced by doing so. There exist many elliptically fibered CY threefolds that give rise to $\mathfrak{u(1)}$s that are non-enhanceable in the sense that they cannot be enhanced without inducing a codimension-one $(4,6,12)$ singularity \cite{MPT}. The possibility of constructing Calabi-Yau manifolds by non-minimal transformations suggests that there might exist models with $\mathfrak{u(1)}$ gauge symmetry that are non-enhanceable for a different reason---because they simply do not have moduli that can be tuned to enhance the $\mathfrak{u(1)}$ into anything else.

Let us explain this point. Suppose there exists some Weierstrass model \eq{MP} that can be mapped to a Calabi-Yau manifold by the minimal transition \eq{map}. In order for such a map to be possible, i.e., in order for the equations \eq{relation} to hold for some $f$ and $g$, the $b$ and $c_i$ of equation \eq{MP} must be tuned. It is imaginable that perhaps for some $D$, $b$ may be completely fixed in order for there to exist a section $a \in \cO(D)$ such that
\be
a^4  \Big|
\left( c_1 c_3-b^2 c_0 -{c_2^2 \ov 3}\right) \,, \quad
a^6 \Big| \left( c_0 c_3^2 -{1 \ov 3} c_1 c_2 c_3 +{2 \ov 27} c_2^3
-{2 \ov 3} b^2 c_0 c_2\ +{b^2 c_1^2 \ov 4} \right) \,.
\ee
If this were the case, it would be impossible to enhance the $\mathfrak{u(1)}$ gauge symmetry further as $b$ cannot be taken to zero. It would be extremely interesting to see if this scenario can be realized.

\acknowledgments{We would like to thank Wati Taylor for discussions and comments on the draft. DRM thanks the Institut Henri Poincar\'e for hospitality. The work of DRM is supported in part by National Science Foundation grant PHY-1307513 (USA) and by the Centre National de la Recherche Scientifique (France). The work of DSP is supported by DOE grant DOE-SC0010008.}

\appendix

\section{Explicit expressions}\label{ap:expressions}

Let us first collect some explicit expressions needed to describe the Klevers-Taylor models. The expressions for $f$ and $g$ can be found in the original paper \cite{KT}. $a_i$ and $b$ are given in terms of $s_i$ as the following:
\bea
a_1 &= {1 \ov 4}(3 s_4 s_6^2 s_8^3 - 4 s_3 s_6 s_7 s_8^3 + 4 s_2 s_7^2 s_8^3 -  12 s_4 s_5 s_6 s_8^2 s_9 + s_3 s_6^2 s_8^2 s_9 + 8 s_3 s_5 s_7 s_8^2 s_9\\
&\qquad- 12 s_1 s_7^2 s_8^2 s_9 + 12 s_4 s_5^2 s_8 s_9^2 - s_2 s_6^2 s_8 s_9^2 - 8 s_2 s_5 s_7 s_8 s_9^2 + 12 s_1 s_6 s_7 s_8 s_9^2\\
&\qquad- 4 s_3 s_5^2 s_9^3 + 4 s_2 s_5 s_6 s_9^3 - 3 s_1 s_6^2 s_9^3) \\
a_2 &= {1 \ov 8} (-s_6^2 s_7 s_8^2 + 4 s_5 s_7^2 s_8^2 + 12 s_4 s_6 s_8^3 - 8 s_3 s_7 s_8^3 + 
   s_6^3 s_8 s_9 - 4 s_5 s_6 s_7 s_8 s_9 - 24 s_4 s_5 s_8^2 s_9 \\
&\qquad-4 s_3 s_6 s_8^2 s_9 + 16 s_2 s_7 s_8^2 s_9 - s_5 s_6^2 s_9^2 +4 s_5^2 s_7 s_9^2 + 16 s_3 s_5 s_8 s_9^2 - 4 s_2 s_6 s_8 s_9^2\\
&\qquad-24 s_1 s_7 s_8 s_9^2 - 8 s_2 s_5 s_9^3 + 12 s_1 s_6 s_9^3)\\
a_3 &= s_4 s_8^3 - s_3 s_8^2 s_9 + s_2 s_8 s_9^2 - s_1 s_9^3 \\
b&=s_7 s_8^2 - s_6 s_8 s_9 + s_5 s_9^2 \,.
\eea

As noted in section \ref{s:result}, $(a_1 a_3 - a_2^2/3)$ is divisible by $b$. The quotient, which we denote $L$, is given by
\bea\label{large L}
L = {1 \ov 192}& (-s_6^4 s_7 s_8^2 + 8 s_5 s_6^2 s_7^2 s_8^2 - 16 s_5^2 s_7^3 s_8^2 + 
   24 s_4 s_6^3 s_8^3 - 96 s_4 s_5 s_6 s_7 s_8^3 - 16 s_3 s_6^2 s_7 s_8^3\\
& +64 s_3 s_5 s_7^2 s_8^3 - 64 s_3^2 s_7 s_8^4 + 192 s_2 s_4 s_7 s_8^4 + 
   s_6^5 s_8 s_9 - 8 s_5 s_6^3 s_7 s_8 s_9 + 16 s_5^2 s_6 s_7^2 s_8 s_9\\
&-48 s_4 s_5 s_6^2 s_8^2 s_9 - 8 s_3 s_6^3 s_8^2 s_9 + 
   192 s_4 s_5^2 s_7 s_8^2 s_9 + 32 s_3 s_5 s_6 s_7 s_8^2 s_9 + 
   32 s_2 s_6^2 s_7 s_8^2 s_9\\
&- 128 s_2 s_5 s_7^2 s_8^2 s_9 + 
   64 s_3^2 s_6 s_8^3 s_9 - 192 s_2 s_4 s_6 s_8^3 s_9 + 64 s_2 s_3 s_7 s_8^3 s_9 \\
&-   576 s_1 s_4 s_7 s_8^3 s_9 - s_5 s_6^4 s_9^2 + 8 s_5^2 s_6^2 s_7 s_9^2 - 
   16 s_5^3 s_7^2 s_9^2 + 32 s_3 s_5 s_6^2 s_8 s_9^2 - 8 s_2 s_6^3 s_8 s_9^2 \\
&-128 s_3 s_5^2 s_7 s_8 s_9^2 + 32 s_2 s_5 s_6 s_7 s_8 s_9^2 - 
   48 s_1 s_6^2 s_7 s_8 s_9^2 + 192 s_1 s_5 s_7^2 s_8 s_9^2\\
&-64 s_3^2 s_5 s_8^2 s_9^2 + 192 s_2 s_4 s_5 s_8^2 s_9^2 - 
   64 s_2 s_3 s_6 s_8^2 s_9^2 + 576 s_1 s_4 s_6 s_8^2 s_9^2 - 
   64 s_2^2 s_7 s_8^2 s_9^2\\
&+ 192 s_1 s_3 s_7 s_8^2 s_9^2 - 
   16 s_2 s_5 s_6^2 s_9^3 + 24 s_1 s_6^3 s_9^3 + 64 s_2 s_5^2 s_7 s_9^3 - 
   96 s_1 s_5 s_6 s_7 s_9^3\\
&+ 64 s_2 s_3 s_5 s_8 s_9^3 - 576 s_1 s_4 s_5 s_8 s_9^3 + 
   64 s_2^2 s_6 s_8 s_9^3 - 192 s_1 s_3 s_6 s_8 s_9^3 - 64 s_2^2 s_5 s_9^4\\
&+ 
   192 s_1 s_3 s_5 s_9^4)
\eea

We now collect some polynomials of $s_i$ required to describe the Weierstrass model $\cX'$ birationally equivalent to the KT models. When $d$ and $k$ are given by equations \eq{d} and \eq{k}, $c_0$ is given by
\bea\label{c0}
c_0 =
{1 \ov 4} &(s_4^4 s_5^2 s_8^8 - 4 s_1 s_4^4 s_8^9 - 4 s_3 s_4^3 s_5^2 s_8^7 s_9 + 
   16 s_1 s_3 s_4^3 s_8^8 s_9 + 6 s_3^2 s_4^2 s_5^2 s_8^6 s_9^2 \\& + 
   4 s_2 s_4^3 s_5^2 s_8^6 s_9^2 - 2 s_1 s_4^3 s_5 s_6 s_8^6 s_9^2 - 
   24 s_1 s_3^2 s_4^2 s_8^7 s_9^2 - 12 s_1 s_2 s_4^3 s_8^7 s_9^2 - 
   4 s_3^3 s_4 s_5^2 s_8^5 s_9^3 \\& - 12 s_2 s_3 s_4^2 s_5^2 s_8^5 s_9^3 + 
   6 s_1 s_3 s_4^2 s_5 s_6 s_8^5 s_9^3 + 16 s_1 s_3^3 s_4 s_8^6 s_9^3 + 
   36 s_1 s_2 s_3 s_4^2 s_8^6 s_9^3 + 4 s_1^2 s_4^3 s_8^6 s_9^3 \\& + 
   s_3^4 s_5^2 s_8^4 s_9^4 + 12 s_2 s_3^2 s_4 s_5^2 s_8^4 s_9^4 + 
   6 s_2^2 s_4^2 s_5^2 s_8^4 s_9^4 - 6 s_1 s_3^2 s_4 s_5 s_6 s_8^4 s_9^4 - 
   6 s_1 s_2 s_4^2 s_5 s_6 s_8^4 s_9^4 \\&+ s_1^2 s_4^2 s_6^2 s_8^4 s_9^4 + 
   2 s_1^2 s_4^2 s_5 s_7 s_8^4 s_9^4 - 4 s_1 s_3^4 s_8^5 s_9^4 - 
   36 s_1 s_2 s_3^2 s_4 s_8^5 s_9^4 - 12 s_1 s_2^2 s_4^2 s_8^5 s_9^4 \\&- 
   16 s_1^2 s_3 s_4^2 s_8^5 s_9^4 - 4 s_2 s_3^3 s_5^2 s_8^3 s_9^5 - 
   12 s_2^2 s_3 s_4 s_5^2 s_8^3 s_9^5 + 2 s_1 s_3^3 s_5 s_6 s_8^3 s_9^5 + 
   12 s_1 s_2 s_3 s_4 s_5 s_6 s_8^3 s_9^5 \\&- 2 s_1^2 s_3 s_4 s_6^2 s_8^3 s_9^5 - 
   4 s_1^2 s_3 s_4 s_5 s_7 s_8^3 s_9^5 + 12 s_1 s_2 s_3^3 s_8^4 s_9^5 + 
   24 s_1 s_2^2 s_3 s_4 s_8^4 s_9^5 + 20 s_1^2 s_3^2 s_4 s_8^4 s_9^5 \\&+ 
   8 s_1^2 s_2 s_4^2 s_8^4 s_9^5 + 6 s_2^2 s_3^2 s_5^2 s_8^2 s_9^6 + 
   4 s_2^3 s_4 s_5^2 s_8^2 s_9^6 - 6 s_1 s_2 s_3^2 s_5 s_6 s_8^2 s_9^6 - 
   6 s_1 s_2^2 s_4 s_5 s_6 s_8^2 s_9^6 \\&+ s_1^2 s_3^2 s_6^2 s_8^2 s_9^6 + 
   2 s_1^2 s_2 s_4 s_6^2 s_8^2 s_9^6 + 2 s_1^2 s_3^2 s_5 s_7 s_8^2 s_9^6+ 
   4 s_1^2 s_2 s_4 s_5 s_7 s_8^2 s_9^6 - 2 s_1^3 s_4 s_6 s_7 s_8^2 s_9^6\\& - 
   12 s_1 s_2^2 s_3^2 s_8^3 s_9^6 - 8 s_1^2 s_3^3 s_8^3 s_9^6 - 
   4 s_1 s_2^3 s_4 s_8^3 s_9^6 - 24 s_1^2 s_2 s_3 s_4 s_8^3 s_9^6 + 
   4 s_1^3 s_4^2 s_8^3 s_9^6 \\&- 4 s_2^3 s_3 s_5^2 s_8 s_9^7 + 
   6 s_1 s_2^2 s_3 s_5 s_6 s_8 s_9^7 - 2 s_1^2 s_2 s_3 s_6^2 s_8 s_9^7 - 
   4 s_1^2 s_2 s_3 s_5 s_7 s_8 s_9^7 + 2 s_1^3 s_3 s_6 s_7 s_8 s_9^7 \\&+ 
   4 s_1 s_2^3 s_3 s_8^2 s_9^7 + 16 s_1^2 s_2 s_3^2 s_8^2 s_9^7 + 
   4 s_1^2 s_2^2 s_4 s_8^2 s_9^7 + s_2^4 s_5^2 s_9^8 - 2 s_1 s_2^3 s_5 s_6 s_9^8 + 
   s_1^2 s_2^2 s_6^2 s_9^8 \\&+ 2 s_1^2 s_2^2 s_5 s_7 s_9^8 - 
   2 s_1^3 s_2 s_6 s_7 s_9^8 + s_1^4 s_7^2 s_9^8 - 8 s_1^2 s_2^2 s_3 s_8 s_9^8 - 
   4 s_1^3 s_3^2 s_8 s_9^8 \\&+ 4 s_1^3 s_2 s_4 s_8 s_9^8 + 4 s_1^3 s_2 s_3 s_9^9 - 
   4 s_1^4 s_4 s_9^9) \,.
\eea
$c_1$ is given by
\bea\label{c1}
c_1 = {1 \ov 2}& (-s_4^3 s_5 s_6 s_8^7 + 2 s_2 s_4^3 s_8^8 + 2 s_4^3 s_5^2 s_8^6 s_9
  +3 s_3 s_4^2 s_5 s_6 s_8^6 s_9  
- 6 s_2 s_3 s_4^2 s_8^7 s_9 - 
   6 s_1 s_4^3 s_8^7 s_9 \\
&   - 6 s_3 s_4^2 s_5^2 s_8^5 s_9^2 - 
   3 s_3^2 s_4 s_5 s_6 s_8^5 s_9^2 
- 3 s_2 s_4^2 s_5 s_6 s_8^5 s_9^2 + 
   s_1 s_4^2 s_6^2 s_8^5 s_9^2 + 2 s_1 s_4^2 s_5 s_7 s_8^5 s_9^2 \\
&+ 
   6 s_2 s_3^2 s_4 s_8^6 s_9^2 
+ 6 s_2^2 s_4^2 s_8^6 s_9^2 + 
   14 s_1 s_3 s_4^2 s_8^6 s_9^2 + 6 s_3^2 s_4 s_5^2 s_8^4 s_9^3 + 
   6 s_2 s_4^2 s_5^2 s_8^4 s_9^3 \\
&+ s_3^3 s_5 s_6 s_8^4 s_9^3 + 
   6 s_2 s_3 s_4 s_5 s_6 s_8^4 s_9^3 - 3 s_1 s_4^2 s_5 s_6 s_8^4 s_9^3 
-  2 s_1 s_3 s_4 s_6^2 s_8^4 s_9^3 - 4 s_1 s_3 s_4 s_5 s_7 s_8^4 s_9^3\\ 
& -
   2 s_2 s_3^3 s_8^5 s_9^3 - 12 s_2^2 s_3 s_4 s_8^5 s_9^3 
 - 
   10 s_1 s_3^2 s_4 s_8^5 s_9^3 - 16 s_1 s_2 s_4^2 s_8^5 s_9^3 - 
   2 s_3^3 s_5^2 s_8^3 s_9^4\\
& - 12 s_2 s_3 s_4 s_5^2 s_8^3 s_9^4 
- 
   3 s_2 s_3^2 s_5 s_6 s_8^3 s_9^4 - 3 s_2^2 s_4 s_5 s_6 s_8^3 s_9^4 + 
   6 s_1 s_3 s_4 s_5 s_6 s_8^3 s_9^4 + s_1 s_3^2 s_6^2 s_8^3 s_9^4 \\
&+ 
   2 s_1 s_2 s_4 s_6^2 s_8^3 s_9^4 + 2 s_1 s_3^2 s_5 s_7 s_8^3 s_9^4 + 
   4 s_1 s_2 s_4 s_5 s_7 s_8^3 s_9^4 - 3 s_1^2 s_4 s_6 s_7 s_8^3 s_9^4
+ 
   6 s_2^2 s_3^2 s_8^4 s_9^4 \\
& + 2 s_1 s_3^3 s_8^4 s_9^4 
+ 
   6 s_2^3 s_4 s_8^4 s_9^4 + 24 s_1 s_2 s_3 s_4 s_8^4 s_9^4
+ 
   12 s_1^2 s_4^2 s_8^4 s_9^4 + 6 s_2 s_3^2 s_5^2 s_8^2 s_9^5  \\
& + 
   6 s_2^2 s_4 s_5^2 s_8^2 s_9^5 + 3 s_2^2 s_3 s_5 s_6 s_8^2 s_9^5 
- 
   3 s_1 s_3^2 s_5 s_6 s_8^2 s_9^5 - 6 s_1 s_2 s_4 s_5 s_6 s_8^2 s_9^5 - 
   2 s_1 s_2 s_3 s_6^2 s_8^2 s_9^5 \\
&+ s_1^2 s_4 s_6^2 s_8^2 s_9^5 
- 
   4 s_1 s_2 s_3 s_5 s_7 s_8^2 s_9^5 + 2 s_1^2 s_4 s_5 s_7 s_8^2 s_9^5 + 
   3 s_1^2 s_3 s_6 s_7 s_8^2 s_9^5  - 6 s_2^3 s_3 s_8^3 s_9^5  \\
&- 
   8 s_1 s_2 s_3^2 s_8^3 s_9^5
- 14 s_1 s_2^2 s_4 s_8^3 s_9^5 - 
   16 s_1^2 s_3 s_4 s_8^3 s_9^5 - 6 s_2^2 s_3 s_5^2 s_8 s_9^6 
- 
   s_2^3 s_5 s_6 s_8 s_9^6 \\
&+ 6 s_1 s_2 s_3 s_5 s_6 s_8 s_9^6 + 
   s_1 s_2^2 s_6^2 s_8 s_9^6 - s_1^2 s_3 s_6^2 s_8 s_9^6 
+ 
   2 s_1 s_2^2 s_5 s_7 s_8 s_9^6 - 2 s_1^2 s_3 s_5 s_7 s_8 s_9^6 \\
& - 
   3 s_1^2 s_2 s_6 s_7 s_8 s_9^6 + 2 s_1^3 s_7^2 s_8 s_9^6
+ 2 s_2^4 s_8^2 s_9^6 + 
   10 s_1 s_2^2 s_3 s_8^2 s_9^6 + 4 s_1^2 s_3^2 s_8^2 s_9^6 \\
&+ 
   14 s_1^2 s_2 s_4 s_8^2 s_9^6 + 2 s_2^3 s_5^2 s_9^7 - 3 s_1 s_2^2 s_5 s_6 s_9^7 +
    s_1^2 s_2 s_6^2 s_9^7 
+ 2 s_1^2 s_2 s_5 s_7 s_9^7 \\
& - s_1^3 s_6 s_7 s_9^7 - 
   4 s_1 s_2^3 s_8 s_9^7 - 6 s_1^2 s_2 s_3 s_8 s_9^7 - 6 s_1^3 s_4 s_8 s_9^7 + 
   2 s_1^2 s_2^2 s_9^8 + 2 s_1^3 s_3 s_9^8)
\eea

\bibliographystyle{JHEP}
\bibliography{NMU1}

\providecommand{\href}[2]{#2}\begingroup\raggedright\begin{thebibliography}{10}

\bibitem{Vafa}
C.~Vafa, {\it {Evidence for F theory}},  {\em Nucl.Phys.} {\bf B469} (1996)
  403--418, [\href{http://arxiv.org/abs/hep-th/9602022}{{\tt hep-th/9602022}}].

\bibitem{MV1}
D.~R. Morrison and C.~Vafa, {\it {Compactifications of F theory on Calabi-Yau
  threefolds. 1}},  {\em Nucl.Phys.} {\bf B473} (1996) 74--92,
  [\href{http://arxiv.org/abs/hep-th/9602114}{{\tt hep-th/9602114}}].

\bibitem{MV2}
D.~R. Morrison and C.~Vafa, {\it {Compactifications of F theory on Calabi-Yau
  threefolds. 2.}},  {\em Nucl.Phys.} {\bf B476} (1996) 437--469,
  [\href{http://arxiv.org/abs/hep-th/9603161}{{\tt hep-th/9603161}}].

\bibitem{KT}
D.~Klevers and W.~Taylor, {\it {Three-Index Symmetric Matter Representations of
  SU(2) in F-Theory from Non-Tate Form Weierstrass Models}},
  \href{http://arxiv.org/abs/1604.01030}{{\tt arXiv:1604.01030}}.

\bibitem{Klevers:2014bqa}
D.~Klevers, D.~K. Mayorga~Pena, P.-K. Oehlmann, H.~Piragua, and J.~Reuter, {\it
  {F-Theory on all Toric Hypersurface Fibrations and its Higgs Branches}},
  {\em JHEP} {\bf 01} (2015) 142, [\href{http://arxiv.org/abs/1408.4808}{{\tt
  arXiv:1408.4808}}].

\bibitem{MP}
D.~R. Morrison and D.~S. Park, {\it {F-Theory and the Mordell-Weil Group of
  Elliptically-Fibered Calabi-Yau Threefolds}},  {\em JHEP} {\bf 10} (2012)
  128, [\href{http://arxiv.org/abs/1208.2695}{{\tt arXiv:1208.2695}}].

\bibitem{Mayrhofer:2012zy}
C.~Mayrhofer, E.~Palti, and T.~Weigand, {\it {U(1) symmetries in F-theory GUTs
  with multiple sections}},  {\em JHEP} {\bf 03} (2013) 098,
  [\href{http://arxiv.org/abs/1211.6742}{{\tt arXiv:1211.6742}}].

\bibitem{Braun:2013yti}
V.~Braun, T.~W. Grimm, and J.~Keitel, {\it {New Global F-theory GUTs with U(1)
  symmetries}},  {\em JHEP} {\bf 09} (2013) 154,
  [\href{http://arxiv.org/abs/1302.1854}{{\tt arXiv:1302.1854}}].

\bibitem{Cvetic:2013nia}
M.~Cvetic, D.~Klevers, and H.~Piragua, {\it {F-Theory Compactifications with
  Multiple U(1)-Factors: Constructing Elliptic Fibrations with Rational
  Sections}},  {\em JHEP} {\bf 06} (2013) 067,
  [\href{http://arxiv.org/abs/1303.6970}{{\tt arXiv:1303.6970}}].

\bibitem{Braun:2013nqa}
V.~Braun, T.~W. Grimm, and J.~Keitel, {\it {Geometric Engineering in Toric
  F-Theory and GUTs with U(1) Gauge Factors}},  {\em JHEP} {\bf 12} (2013) 069,
  [\href{http://arxiv.org/abs/1306.0577}{{\tt arXiv:1306.0577}}].

\bibitem{Borchmann:2013hta}
J.~Borchmann, C.~Mayrhofer, E.~Palti, and T.~Weigand, {\it {SU(5) Tops with
  Multiple U(1)s in F-theory}},  {\em Nucl. Phys.} {\bf B882} (2014) 1--69,
  [\href{http://arxiv.org/abs/1307.2902}{{\tt arXiv:1307.2902}}].

\bibitem{Antoniadis:2014jma}
I.~Antoniadis and G.~K. Leontaris, {\it {F-GUTs with Mordell–Weil U(1) 's}},
  {\em Phys. Lett.} {\bf B735} (2014) 226--230,
  [\href{http://arxiv.org/abs/1404.6720}{{\tt arXiv:1404.6720}}].

\bibitem{Kuntzler:2014ila}
M.~Kuntzler and S.~Schafer-Nameki, {\it {Tate Trees for Elliptic Fibrations
  with Rank one Mordell-Weil group}},
  \href{http://arxiv.org/abs/1406.5174}{{\tt arXiv:1406.5174}}.

\bibitem{Esole:2014dea}
M.~Esole, M.~J. Kang, and S.-T. Yau, {\it {A New Model for Elliptic Fibrations
  with a Rank One Mordell-Weil Group: I. Singular Fibers and Semi-Stable
  Degenerations}},  \href{http://arxiv.org/abs/1410.0003}{{\tt
  arXiv:1410.0003}}.

\bibitem{Lawrie:2015hia}
C.~Lawrie, S.~Schafer-Nameki, and J.-M. Wong, {\it {F-theory and All Things
  Rational: Surveying U(1) Symmetries with Rational Sections}},  {\em JHEP}
  {\bf 09} (2015) 144, [\href{http://arxiv.org/abs/1504.05593}{{\tt
  arXiv:1504.05593}}].

\bibitem{Krippendorf:2015kta}
S.~Krippendorf, S.~Schafer-Nameki, and J.-M. Wong, {\it {Froggatt-Nielsen meets
  Mordell-Weil: A Phenomenological Survey of Global F-theory GUTs with U(1)s}},
   {\em JHEP} {\bf 11} (2015) 008, [\href{http://arxiv.org/abs/1507.05961}{{\tt
  arXiv:1507.05961}}].

\bibitem{MR1078016}
R.~Miranda, {\em The basic theory of elliptic surfaces}.
\newblock ETS Editrice, Pisa, 1989.

\bibitem{instK3}
P.~S. Aspinwall and D.~R. Morrison, {\it Point-like instantons on {K3}
  orbifolds},  {\em Nuclear Phys. B} {\bf 503} (1997) 533--564,
  [\href{http://arxiv.org/abs/hep-th/9705104}{{\tt hep-th/9705104}}].

\bibitem{nongeometries}
J.~McOrist, D.~R. Morrison, and S.~Sethi, {\it Geometries, non-geometries, and
  fluxes},  {\em Adv. Theor. Math. Phys.} {\bf 14} (2010) 1515--1583,
  [\href{http://arxiv.org/abs/1004.5447}{{\tt arXiv:1004.5447}}].

\bibitem{Shioda1}
T.~Shioda, {\it Mordell-{W}eil lattices and {G}alois representation. {I}},
  {\em Proc. Japan Acad. Ser. A Math. Sci.} {\bf 65} (1989), no.~7 268--271.

\bibitem{Shioda2}
T.~Shioda, {\it On the {M}ordell-{W}eil lattices},  {\em Comment. Math. Univ.
  St. Paul.} {\bf 39} (1990), no.~2 211--240.

\bibitem{Intersection}
D.~S. Park, {\it {Anomaly Equations and Intersection Theory}},  {\em JHEP} {\bf
  01} (2012) 093, [\href{http://arxiv.org/abs/1111.2351}{{\tt
  arXiv:1111.2351}}].

\bibitem{Cvetic:2012xn}
M.~Cvetic, T.~W. Grimm, and D.~Klevers, {\it {Anomaly Cancellation And Abelian
  Gauge Symmetries In F-theory}},  {\em JHEP} {\bf 02} (2013) 101,
  [\href{http://arxiv.org/abs/1210.6034}{{\tt arXiv:1210.6034}}].

\bibitem{Cvetic:2015ioa}
M.~Cvetic, D.~Klevers, H.~Piragua, and W.~Taylor, {\it {General U(1)xU(1)
  F-theory compactifications and beyond: geometry of unHiggsings and novel
  matter structure}},  {\em JHEP} {\bf 11} (2015) 204,
  [\href{http://arxiv.org/abs/1507.05954}{{\tt arXiv:1507.05954}}].

\bibitem{Morrison:2014era}
D.~R. Morrison and W.~Taylor, {\it {Sections, multisections, and U(1) fields in
  F-theory}},  \href{http://arxiv.org/abs/1404.1527}{{\tt arXiv:1404.1527}}.

\bibitem{MPT}
D.~R. Morrison, D.~S. Park, and W.~Taylor. {to appear}.

\end{thebibliography}\endgroup

\end{document}